\documentclass[10pt,english,reprint,superscriptaddress,aps,prl,floatfix]{revtex4-2}
\usepackage[T1]{fontenc}
\usepackage[utf8]{inputenc}
\usepackage{xcolor}
\usepackage{babel}
\usepackage{amsmath}
\usepackage{amsthm}
\usepackage{graphicx}
\usepackage[pdfusetitle,
 bookmarks=true,bookmarksnumbered=false,bookmarksopen=false,
 breaklinks=false,pdfborder={0 0 1},backref=false,colorlinks=true]
 {hyperref}
\hypersetup{
 linkcolor=blue,citecolor=blue,urlcolor=blue}

\makeatletter
\usepackage{babel}
\usepackage{color}
\usepackage[normalem]{ulem}
\usepackage{amsthm}
\usepackage{xcolor}

\makeatother

\begin{document}

\title{Dijet hadrochemistry: tomography of jet-induced medium excitation and diffusion wake}

\author{Cong Yi}
\email{congyi@ccnu.edu.cn}
\affiliation{Institute of Particle Physics and Key Laboratory of Quark and Lepton
Physics (MOE), Central China Normal University, Wuhan 430079, China}

\author{Xiang-Yu Wu}
\email{xiangyu.wu2@mail.mcgill.ca}
\affiliation{Department of Physics, McGill University, Montreal, QC, Canada H3A
2T8}

\author{Zhong Yang}
\email{zhong.yang@vanderbilt.edu}
\affiliation{Department of Physics and Astronomy, Vanderbilt University, Nashville,
TN 37235}

\author{Pi Duan}
\email{duanpi@mails.ccnu.edu.cn}
\affiliation{Institute of Particle Physics and Key Laboratory of Quark and Lepton
Physics (MOE), Central China Normal University, Wuhan 430079, China}

\author{Guang-You Qin}
\email{guangyou.qin@ccnu.edu.cn}
\affiliation{Institute of Particle Physics and Key Laboratory of Quark and Lepton
Physics (MOE), Central China Normal University, Wuhan 430079, China}

\author{Xin-Nian Wang}
\email{xnwang@lbl.gov}
\affiliation{Institute of Particle Physics and Key Laboratory of Quark and Lepton
Physics (MOE), Central China Normal University, Wuhan 430079, China}

\author{Wenbin Zhao}
\email{WenbinZhao@ccnu.edu.cn}
\affiliation{Institute of Particle Physics and Key Laboratory of Quark and Lepton
Physics (MOE), Central China Normal University, Wuhan 430079, China}

\begin{abstract}
Energetic partons traversing the quark-gluon plasma in relativistic heavy-ion collisions induce a medium response that modifies the production of associated hadrons, whose chemical composition provides a sensitive probe of this response. We study jet-associated hadrochemistry using the coupled linear Boltzmann transport and hydrodynamics model with a hybrid hadronization scheme that combines Cooper--Frye particlization, quark coalescence and string fragmentation. For inclusive charged jets in $p+p$ and 0--10\% central Pb+Pb collisions at $\sqrt{s_{\mathrm{NN}}}=5.02$~TeV, the calculated in-cone $K/\pi$ and $p/\pi$ ratios are enhanced in Pb+Pb relative to $p+p$ collisions, consistent with preliminary ALICE measurements, with coalescence between shower and thermal partons providing a substantial contribution. Extending the analysis to dijet events, we predict that both ratios are enhanced around the leading and subleading jets, more strongly around the subleading jet and more pronounced for larger radial size. After mixed-event acceptance correction and sideband background subtraction, the yields of associated pions, kaons and protons all become negative near $|\Delta\phi|\simeq\pi/2$, signaling a depletion caused by the diffusion wake. The $K/\pi$ and $p/\pi$ ratios of these deficits remain finite and exceed the $p+p$ values around the jet peaks, indicating that the missing hadrons originate from jet-excited medium. These results establish dijet hadrochemistry as an experimentally accessible tomographic probe of jet-induced medium excitation and the diffusion wake.
\end{abstract}

\maketitle

\textit{Introduction.---} 
In relativistic heavy-ion collisions at the Relativistic Heavy Ion Collider (RHIC) and the Large Hadron Collider (LHC), energetic partons produced in initial hard scatterings traverse and lose energy in the quark-gluon plasma (QGP), leading to the suppression of high-$p_{\mathrm{T}}$ hadron and jet yields, as well as the modification of jet substructures~\cite{Wang:1992qdg,Braaten:1991we,Gyulassy:1993hr,Baier:1996kr,Zakharov:1996fv,Gyulassy:1999zd,Wiedemann:2000za,Wang:2001ifa,Arnold:2002ja,Qin:2007rn,Djordjevic:2008iz,Bass:2008rv,Majumder:2010qh,Qin:2015srf,Blaizot:2015lma,Cao:2020wlm,Wang:2025lct}.  These phenomena are collectively known as jet quenching, which provides one of the key evidences for the QGP formation~\cite{Wang:1992qdg,Majumder:2010qh,Qin:2015srf,Blaizot:2015lma,Cao:2020wlm}. Over the past decades, jet quenching studies~\cite{Qin:2010mn,Majumder:2011uk,Dai:2012am,Blaizot:2013hx,Kang:2014xsa,Chien:2016led,Caucal:2018dla,Xing:2019xae,Huss:2020dwe,Mehtar-Tani:2021fud,Zhao:2021vmu,Cao:2024pxc,Xing:2024yrb,Andres:2024xvk,Schenke:2009gb,Zapp:2012ak,Casalderrey-Solana:2014bpa,Cao:2017qpx,JETSCAPE:2017eso,Luo:2023nsi,Karpenko:2024fgg,Modarresi-Yazdi:2024vfh,Yazdi:2022bru,Schenke:2009gb,Shi:2022rja,Young:2011qx} have placed quantitative constraints on the transport properties of the QGP, most notably the jet quenching parameter $\hat{q}$~\cite{JET:2013cls,JETSCAPE:2021ehl,Xie:2022ght,Chen:2024epd,Karmakar:2024jak,Karmakar:2023ity}, which quantifies the transverse momentum broadening of partons per unit length and also controls the rate of medium-induced gluon radiation. 

Jet-medium interaction not only modifies jets themselves, but also deposits energy and momentum into the medium and thereby induces a medium response~\cite{He:2018xjv,Cao:2022odi,JETSCAPE:2022jer,He:2022evt}. The classical picture of jet-induced medium response consists of a Mach-cone-like wake front accompanied by a diffusion wake trailing the propagating jet~\cite{Casalderrey-Solana:2004fdk,Stoecker:2004qu,Ruppert:2005uz,Chaudhuri:2005vc,Chakraborty:2006md,Gubser:2007ga,Chesler:2007an,Qin:2009uh,Neufeld:2009ep,Li:2010ts,Bouras:2012mh,Ayala:2016pvm,Yan:2017rku,Casalderrey-Solana:2020rsj,Barata:2024ukm}. Early theoretical studies have shown that the angular structure of jet-induced Mach cone is sensitive to the speed of sound and the viscosity of the QGP~\cite{Neufeld:2008dx,Bouras:2014rea}. However, direct observation of the Mach-cone signal still remains experimentally challenging. Radial flow and density gradients in the expanding QGP distort the angular structure of the medium response. In additon, interference among the excitations generated by multiple shower partons and the event-by-event fluctuations of the dynamical medium substantially smears the signal~\cite{Ma:2010dv,Betz:2010qh,Neufeld:2011yh,Renk:2013pua,Tachibana:2014lja,Floerchinger:2014yqa,Andrade:2014swa,Gao:2016ldo,Brewer:2017fqy,Tachibana:2017syd, Milhano:2017nzm}.

These complications have motivated the search for a more distinctive manifestation of jet-induced medium response on the back side of the jet, namely the diffusion wake, which tends to deplete the soft hadron production in the final state. A dedicated strategy to detect this depletion in $Z/\gamma$+jet events was proposed in Ref.~\cite{Yang:2021qtl}, in which the contribution from multi-parton interactions (MPI) was estimated and subtracted. Later, CMS found evidence for such a depletion in $Z$+jet events~\cite{CMS:2025dua}, whereas the ATLAS photon-jet measurement observed no statistically significant signal within current uncertainties~\cite{ATLAS:2024prm}. More recently, a background-canceling rapidity asymmetry based on dijets with a finite rapidity interval was proposed~\cite{Yang:2025dqu}, which has led to the first significant observation of the diffusion wake by CMS~\cite{CMS:2026mur}.

Jet-induced medium response can also modify the species composition of jet-associated hadrons~\cite{Luo:2021voy,Luo:2024xog}. Recombination of jet-excited medium partons with thermal partons can enhance baryon-to-meson ratios, while the enhanced strangeness abundance in the QGP can increase strange-hadron production around the quenched jets. Such effect is similar to the baryon-to-meson and strangeness enhancements in the bulk medium~\cite{Rafelski:1982pu,Fries:2003vb,Molnar:2003ff,Greco:2003xt,Greco:2003vf,Hwa:2004ng,STAR:2005npq,PHENIX:2006dpn,STAR:2007zea,Feng:2024tmc,Chen:2021rrp,Sirimanna:2022zje}. 
Recent measurements have begun to explore such jet-induced hadrochemistry observables~\cite{Dale-Gau:2023ree,Cantway:2025yao}. However, existing theoretical calculations rely mainly on a multi-phase transport (AMPT) model~\cite{Lin:2004en,Zhang:2005ni}, which includes only elastic scatterings in partonic evolution and employs a simple quark coalescence scheme for hadronization, neglecting medium-induced radiation and fragmentation. A more complete dynamical description is therefore required for quantitative studies.

More interestingly, while previous studies have examined baryon and strangeness enhancements around the quenched jets, the species composition of the depleted yield associated with the diffusion wake remains unexplored. This raises the question of whether jet-associated hadrochemistry can offer a new means of identifying the diffusion-wake signal and probing the QGP properties through jet-induced medium response. In this Letter, we propose dijet hadrochemistry as such a probe. 
To quantitatively show this, we use the state-of-the-art coupled linear Boltzmann transport and hydrodynamics (CoLBT-hydro) model~\cite{Chen:2017zte,Chen:2020tbl,Chen:2021gkj} with hybrid hydro-coalescence-fragmentation hadronization~\cite{Zhao:2020wcd,Zhao:2021vmu,YuanyuanWang:2023vor} in our calculation. 

\begin{figure}[tb]
\centering \includegraphics[width=1.0\columnwidth]{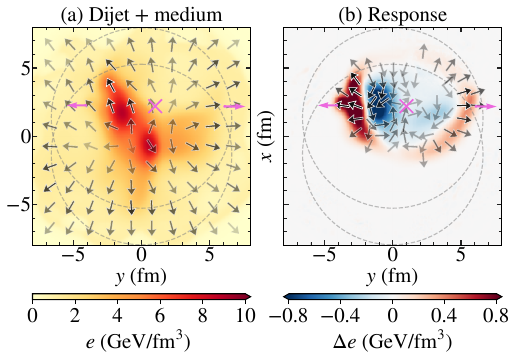}
\caption{Transverse-plane energy density at $\tau = 5.7$~fm/$c$ and $\eta_s = 0$ for (a) the dijet plus medium and (b) the dijet-induced medium response, $\Delta e = e_{\mathrm{dijet+medium}} - e_{\mathrm{bkg}}$, where $e_{\mathrm{bkg}}$ is a reference evolution with identical initial conditions but without the dijet. Black arrows show the local transverse energy-flux direction, with opacity increasing with flux magnitude within each panel. Magenta arrows mark the shower-parton directions, the magenta cross marks the dijet production point, and dashed circles outline the colliding nuclei.}
\label{fig:dijet_wake_tau5p0} 
\end{figure}

Figure~\ref{fig:dijet_wake_tau5p0} shows a typical CoLBT-hydro simulation of dijet events with hydrodynamic medium (model details are provided in the next section), in which the medium responses induced by two jets coexist and interfere. 
Figure~\ref{fig:dijet_wake_tau5p0} (a) shows the fluid energy density in the transverse plane of a representative central Pb+Pb event at $\tau = 5.7$~fm/$c$, with arrows indicating the flow velocity of the bulk medium. Two jets are produced at an off-center point and propagate back-to-back along the $y$-axis through the medium. Subtracting the energy density of a reference evolution with identical initial conditions but without dijet isolates the jet-induced disturbance, $\Delta e = e_{\mathrm{dijet+medium}} - e_{\mathrm{bkg}}$, shown in Fig.~\ref{fig:dijet_wake_tau5p0} (b). The disturbance exhibits two Mach-cone-like fronts of enhanced energy density, each trailed by a depleted diffusion wake. Along the dijet axis, the depletion associated with the leading (subleading) jet, which manifests in the opposite direction, is partially masked by the enhanced front of the subleading (leading) jet~\cite{Yang:2025dqu}. In the regions transverse to the dijet axis, the energy-density depletion left behind by two jets produces a pressure gradient that drives the perturbed energy flux inward, toward the depleted region, and thereby counteracts the outward medium flow. The reduced energy density and weakened outward flow of these cells will suppress hadron production near $|\Delta\phi| \simeq \pi/2$. In this Letter, we are particularly interested in the hadrochemical composition of this depletion caused by the diffusion wake. Our calculation will show that the $K/\pi$ and $p/\pi$ ratios of this depletion exceed the $p+p$ values around the jets, which implies that the missing hadrons originate from jet-excited medium. 

%%%%%%%%%%%%%%%%%%
%%%%%%%%%%%%%%%%%%

\textit{Model and Setup.---} 
In our calculations, the space-time evolution of the QGP and the interaction of energetic partons with the bulk medium are described within the CoLBT-hydro framework~\cite{Chen:2017zte,Chen:2020tbl,Zhao:2021vmu}, which couples linear Boltzmann transport (LBT) model to the event-by-event $(3+1)$-dimensional CCNU-LBNL viscous hydrodynamic model (CLVisc)~\cite{Pang:2012he,Pang:2018zzo,Wu:2021fjf}. Initial jet shower partons are generated with PYTHIA~\cite{Sjostrand:2014zea} and subsequently propagated through the medium with the LBT model, which accounts for both elastic and inelastic scatterings~\cite{Luo:2023nsi}. Parton propagation and hydrodynamics evolution are carried out concurrently, with CLVisc model initialized from TRENTo~\cite{Moreland:2014oya} supplemented by a longitudinal envelope function. At each time step, soft and negative partons from the LBT are assumed to thermalize and incorporated as an source term in hydrodynamic equations, with the soft-parton energy cutoff set to be $2$~GeV in the local rest frame of the fluid. The hydrodynamic parameters are calibrated to identified-hadron $p_{\mathrm{T}}$ spectra and to inclusive $p/\pi$ and $K/\pi$ ratios. Further details of the model setups can be found in Ref.~\cite{Zhao:2021vmu}.

After the QGP freezes out at $T_{f} = 150$MeV, hadronization is treated with a hybrid scheme that combines Cooper--Frye particlization, quark coalescence, and string fragmentation~\cite{Han:2016uhh,Zhao:2020wcd,Zhao:2021vmu}. Since this work focuses on associated hadrons with $p_{\mathrm{T}} < 4$GeV, we explicitly include both shower-thermal parton coalescence with the thermal partons sampled on the freeze-out hypersurface, and shower-shower coalescence. In the coalescence process, we include the excited hadron states following Particle Data Group~\cite{ParticleDataGroup:2026mpi}.  Shower partons that do not undergo coalescence are hadronized with the colorless string scheme in PYTHIA8~\cite{Sjostrand:2014zea}. Hadrons produced by all three mechanisms are then fed into SMASH~\cite{SMASH:2016zqf} for resonance decays.

%%%%%%%%%%%%%%%%%%
\begin{figure}[tb]
\centering \includegraphics[width=1\columnwidth]{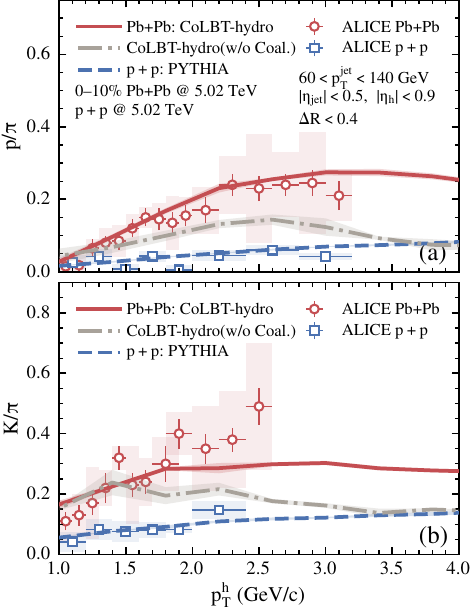}
\caption{Ratios of background-subtracted jet-associated hadron yields, (a) $p/\pi$ and (b) $K/\pi$, within $\Delta R<0.4$ as functions of the associated-hadron $p_{\mathrm{T}}^{h}$. Red solid and gray dash-dotted curves denote the CoLBT-hydro results for 0--10\% Pb+Pb collisions at $5.02$~TeV with and without quark coalescence, respectively. Blue dashed curves show the PYTHIA results for $p+p$ collisions at $5.02$~TeV. 
The kinematic selections are described in the main text, and the preliminary ALICE data are taken from Ref.~\cite{Cantway:2025yao}.
}
\label{fig:single-jet-ratio} 
\end{figure}
%%%%%%%%%%%%%%%%%%

After hadronization, all (charged) hadrons with $p_{\mathrm{T}} > 0.15~\mathrm{GeV}/c$ and $|\eta| < 4$ are selected for full (charged) jet reconstruction. Jets are reconstructed using the anti-$k_{\mathrm{T}}$ algorithm~\cite{Cacciari:2008gp,Cacciari:2011ma}, with a radius parameter $R = 0.4$ to match the LHC measurements~\cite{ATLAS:2018gwx}. Following experimental analysis~\cite{Cantway:2025yao}, we use the SoftKiller method~\cite{Cacciari:2014gra} to subtract the soft underlying-event contamination, with a grid spacing of $a = 0.4$ over $|\eta| < 4$. Our calculation reproduces the inclusive full-jet $p_{\mathrm{T}}$ spectrum measured by ATLAS in 0--10\% central Pb+Pb collisions at $\sqrt{s_{\mathrm{NN}}} = 5.02~\mathrm{TeV}$~\cite{ATLAS:2018gwx}. 
Following Refs.~\cite{CMS:2016qnj,CMS:2026mur}, the jet-hadron correlations are corrected for pair acceptance using the mixed-event method, and for the long-range background (including MPI contributions) using the sideband subtraction method. 
The resulting background-subtracted, per-jet-normalized associated hadron yields are denoted by $Y_i(\Delta\phi, \Delta\eta)$, where $i $ stands for $ \pi^{\pm}$, $K^{\pm}$, and $p/\bar{p}$, respectively. The details about the background subtraction are presented in the Supplemental Materials.

\textit{Inclusive-jet hadrochemistry.---} 
As a benchmark study, we first examine inclusive-jet hadrochemistry through the $p/\pi$ and $K/\pi$ ratios as shown in Fig.~\ref{fig:single-jet-ratio}. Following the ALICE measurements~\cite{Cantway:2025yao}, charged jets are selected with $60 < p_{\mathrm{T}}^{\mathrm{jet}} < 140$~GeV and $|\eta_{\mathrm{jet}}| < 0.5$, and associated hadrons with $|\eta_h| < 0.9$. The ratios are constructed from the background-subtracted yields $Y_i(\Delta\phi, \Delta\eta)$ integrated over $\Delta R = \sqrt{(\Delta\eta)^2 + (\Delta\phi)^2} < 0.4$ in each hadron $p_{\mathrm{T}}^h$ bin. The red solid and blue dashed curves show the CoLBT-hydro results for 0--10\% Pb+Pb collisions and the PYTHIA results for $p+p$ collisions at $\sqrt{s_{\mathrm{NN}}} = 5.02$~TeV. Both ratios are enhanced in Pb+Pb relative to $p+p$ collisions over $1 < p_{\mathrm{T}}^h < 4$ GeV, and the magnitude and momentum dependence of such enhancements are in good agreement with the preliminary ALICE data~\cite{Cantway:2025yao}. When quark coalescence is switched off, the calculated ratios are still enhanced relative to $p+p$ results at low $p_{\mathrm{T}}^h$, but are substantially less than the full calculation and approach the $p+p$ results at high $p_{\mathrm{T}}^h$. This comparison demonstrates that quark coalescence provides an substantial contribution to the baryon and strangeness enhancements inside jets. Coalescence between surviving shower partons and thermal quarks favors baryon over meson production and thereby enhances $p/\pi$, while thermal strange quarks enhance $K/\pi$ through the same shower-thermal mechanism at intermediate $p_{\mathrm{T}}^h$. The in-cone hadrochemistry therefore provides a sensitive probe of jet-induced medium response and the hadronization mechanism of excited medium partons.

%%%%%%%%%%%%%%%%%%
\begin{figure}[tb]
% \centering \includegraphics[width=0.48\columnwidth]{dijet-AA}\hfill{}\includegraphics[width=0.48\columnwidth]{dijet-pp}
\centering \includegraphics[width=1.0\columnwidth]{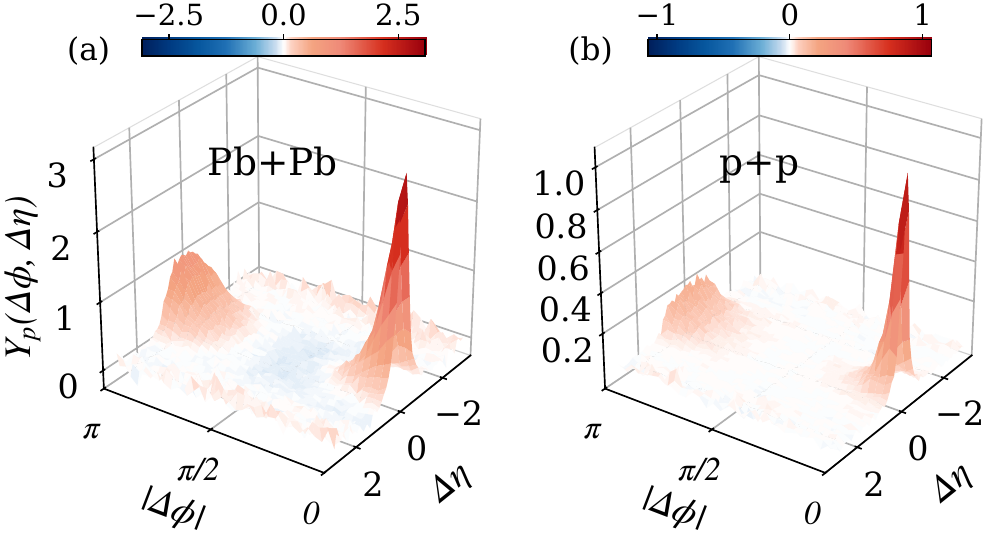}
\caption{
Background-subtracted per-jet associated proton yield $Y_p(\Delta\phi,\Delta\eta)$ 
in (a) $0$--$10\%$ Pb+Pb and (b) $p+p$ collisions at $\sqrt{s_{NN}}=5.02$~TeV.
Associated protons are selected with $1.5<p_T<4$ GeV/$c$. The kinematic selections for dijet events are described in the main text.
}
\label{fig:dijet-2D}
\end{figure}
%%%%%%%%%%%%%%%%%%

\textit{Dijet hadrochemistry and the diffusion wake.---} 
We now extend the analysis to dijet events. To illustrate the interference effect between the medium responses induced by two jets, we calculate the two-dimensional background-subtracted jet-induced hadron yield, e.g., the proton yield $Y_p(\Delta\phi, \Delta\eta)$ shown in Fig.~\ref{fig:dijet-2D}, for Pb+Pb (left) and $p+p$ (right) collisions. Leading and subleading jets are required to have $p_{\mathrm{T}}^{\mathrm{jet1}} > 60$~GeV and $p_{\mathrm{T}}^{\mathrm{jet2}} > 40$GeV, respectively, with $|\eta_{\mathrm{jet1}} - \eta_{\mathrm{jet2}}| < 1$ and $|\Delta\phi_{\mathrm{jet1,jet2}}| > \pi/2$. Relative to the inclusive-jet analysis, the acceptances are extended to $|\eta_{\mathrm{jet}}| < 1.6$ and $|\eta_h| < 2$ to increase the statistics. Associated hadrons are selected within $1.5 < p_{\mathrm{T}}^h < 4$~GeV, a window that retains sensitivity to both the coalescence-driven hadrochemical enhancement at intermediate $p_{\mathrm{T}}^h$ and the soft medium response. The long-range background is estimated from the sideband region $1.5 < |\Delta\eta| < 2.5$ and subtracted with the method described in Ref.~\cite{Cantway:2025yao,CMS:2016qnj}. Since the correlations arising from the anisotropic flow of the bulk medium and from the MPI are approximately independent of $\Delta\eta$ over this range, they are removed by this procedure. Both Pb+Pb and p+p systems exhibit a near-side peak around $(\Delta\eta, \Delta\phi) = (0, 0)$ associated with the leading jet and an away-side structure near $|\Delta\phi| = \pi$ associated with the subleading jet. The structures are broader in Pb+Pb than in $p+p$, reflecting the medium-induced broadening of the jet showers together with the medium-response hadrons carried by the Mach-cone-like wave fronts. Between two jet peaks, the Pb+Pb proton yield falls below the sideband level near $|\Delta\phi| \simeq \pi/2$, resulting in a negative background-subtracted yield, whereas the corresponding $p+p$ yield is consistent with zero. The $p+p$ result thus serves as a closure test of the subtraction procedure in this region, confirming that neither the dijet selection nor the sideband subtraction itself generates a depletion in the absence of a medium. The depletion in Pb+Pb can therefore be attributed to jet-medium interactions and is consistent with the diffusion-wake contribution identified in Fig.~\ref{fig:dijet_wake_tau5p0} (b). Note that the diffusion wake can extend over a broad range in rapidity; part of it may populate the sideband region. Therefore, the subtracted yield, which measures the depletion relative to larger $|\Delta\eta|$, should be regarded as a lower bound on its magnitude.

%%%%%%%%%%%%%%%%%%
\begin{figure}[tb]
\centering \includegraphics[width=1.0\columnwidth]{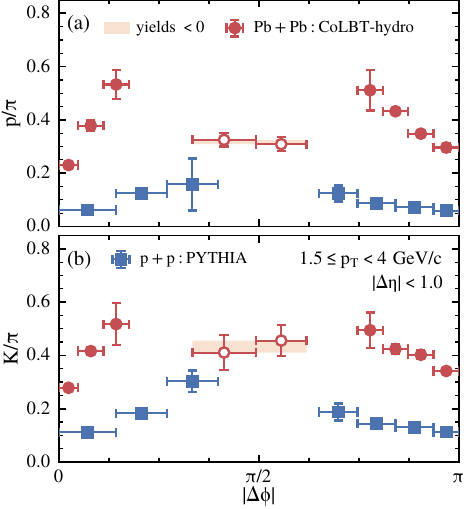}
\caption{Ratios of background-subtracted dijet-associated yields, (a) $p/\pi$ and (b) $K/\pi$, versus $|\Delta\phi|$ relative to the leading-jet axis, integrated over $|\Delta\eta| < 1$. Red circles show CoLBT-hydro results for 0--10\% Pb+Pb collisions and blue squares show PYTHIA results for $p+p$ collisions at 5.02~TeV. Filled symbols denote ratios of positive yields, and open red symbols denote the deficit ratios in bins where all three yields are negative, as indicated by the orange rectangles. The jet and associated-hadron selections are the same as in Fig.~\ref{fig:dijet-2D}.}
\label{fig:dijet-ratio-phi} 
\end{figure}
%%%%%%%%%%%%%%%%%%

We now calculate jet-associated $p/\pi$ and $K/\pi$ ratios as functions of $|\Delta\phi|$ in $p+p$ and central Pb+Pb collisions shown in Fig.~\ref{fig:dijet-ratio-phi}. The ratios are constructed from two-dimensional background-subtracted yields for $1.5 < p_{\mathrm{T}}^h < 4$~GeV, integrated over $|\Delta\eta| < 1$. Detailed procedure is presented in the Supplemental Materials. Only bins in which all three yields share the same sign are retained. In the positive-yield regions near $|\Delta\phi| = 0$ and $\pi$, corresponding to the near-side and away-side jet peaks, both ratios are significantly enhanced in Pb+Pb relative to $p+p$ collisions. In both systems, the ratios rise from jet peaks toward their shoulders, and the rise is considerably steeper in Pb+Pb. This behavior is consistent with a growing contribution from medium response relative to jet fragmentation at larger angles from jet axes. Hadrons at these angles originate predominantly from softer shower and recoil partons, which coalesce more readily with the thermal partons, favoring the production of protons through three-quark recombination and of kaons through the contribution of thermal strange quarks. In $p+p$ collisions, the near-side and away-side ratios are comparable, reflecting similar hadronic composition of vacuum fragmentation for two jets. In Pb+Pb, by contrast, the away-side ratios exceed the near-side values. This asymmetry arises naturally from the dijet selection: compared to the leading jet, the subleading jet typically propagates over longer path, loses more energy, and thus induces stronger medium response. 

More importantly, finite depletion ratios emerge in Pb+Pb collisions around $|\Delta\phi| \simeq \pi/2$, where $Y_p$, $Y_K$, and $Y_\pi$ are all negative. The open symbols in Fig.~\ref{fig:dijet-ratio-phi} denote $|Y_p|/|Y_\pi|$ and $|Y_K|/|Y_\pi|$ ratios, which quantify the proton and kaon deficits relative to the pion deficit. Both depletion ratios are comparable in magnitude to those in the positive-yield regions and well above the corresponding $p+p$ values around jet peaks. This comparison carries a diagnostic value beyond the mere existence of a negative yield. If the depletion were an artifact of over-subtracting jet-fragmentation hadrons, the missing yield would be expected to carry a vacuum-like composition close to the $p+p$ ratios. The enhanced ratios of the deficits instead indicate that the missing hadrons are those that would have been emitted from jet-excited medium. In the $p+p$ reference, no such depletion appears: the background-subtracted yields around $|\Delta\phi| \simeq \pi/2$ are consistent with zero, and the corresponding ratios cannot be reliably determined. We therefore propose the finite and enhanced depletion ratios $|Y_p|/|Y_\pi|$ and $|Y_K|/|Y_\pi|$ as a new hadrochemical signature of the diffusion wake. Unlike the negative yield alone, which is sensitive to the normalization of the background subtraction, these ratios characterize the species composition of the missing yield and are therefore expected to be less sensitive to the overall subtraction uncertainty. Future measurements of these ratios, together with statistically significant negative yields for all three species, would provide a stringent test of this prediction and offer a new means of probing QGP properties through jet-induced medium response.

%%%%%%%%%%%%%%%%%%
\begin{figure}[tb]
\centering \includegraphics[width=1\columnwidth]{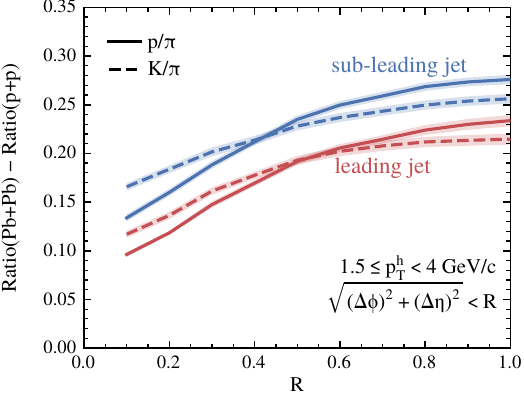}
\caption{Differences between central Pb+Pb and $p+p$ collisions at $\sqrt{s_{NN}} = 5.02$~TeV in the $p/\pi$ and $K/\pi$ ratios constructed from background-subtracted yields within the jet axis, as functions of the cone radius $R$. The solid and dashed curves show differences of $p/\pi$ and $K/\pi$, respectively, around leading (red) and sub-leading (blue) jets. The jet and associated-hadron selections are the same as in Fig.~\ref{fig:dijet-2D}.}
\label{fig:dejet-ratio-R} 
\end{figure}
%%%%%%%%%%%%%%%%%%

Finally, we examine the cone-size $R$ dependence of the hadrochemical modification around the leading and subleading jets. Figure~\ref{fig:dejet-ratio-R} shows the ratio differences between Pb+Pb and $p+p$ collisions, $\Delta\mathcal{R}_i(R) = \mathcal{R}_i^{\mathrm{Pb+Pb}}(R) - \mathcal{R}_i^{p+p}(R)$, where $\mathcal{R}_i(R) = N_i(R)/N_\pi(R)$ for $i = p, K$, and $N_i(R) = \int_{\Delta R < R} Y_i(\Delta\phi, \Delta\eta)\, d\Delta\phi\, d\Delta\eta$ is the background-subtracted yield within a cone of radius $R$ around the corresponding jet axis. Both $p/\pi$ and $K/\pi$ ratios are more  enhanced around the subleading jet, which on average traverses a longer path in the medium and deposits more energy into the medium. The enhancements increase with $R$ because hadrons from medium response dominate over jet fragmentation hadrons at larger angles. For both leading and subleading jets, the $K/\pi$ enhancement exceeds the $p/\pi$ enhancement at small $R$, whereas the latter grows more rapidly and becomes larger at large $R$. This ordering can be understood as follows.
In shower-thermal parton coalescence, a meson requires only one thermal parton whereas a baryon requires two. Consequently, the baryon yield scales approximately quadratically with the local thermal-parton density, while the meson yield scales only linearly. Near the jet axis, where shower partons dominate, the kaon enhancement driven by a single thermal strange quark is stronger, while at larger angles, where the thermal partons dominate, quadratic scaling allows the proton enhancement to take over. 
The jet-induced medium flow further reinforces such trend since the blue shift, which grows with hadron mass, tends to move protons into the intermediate-$p_{\mathrm{T}}^h$ window more efficiently than kaons. 
Therefore, the crossing radius provides a measure of the angular scale at which the medium response overtakes the modified jet shower --- a scale that is expected to depend on how rapidly the jet-deposited energy thermalizes and diffuses in the QGP. 
 
\textit{Summary.---}
In this Letter, we have studied jet-associated hadrochemistry in $p+p$ and 0--10\% central Pb+Pb collisions at $\sqrt{s_{\mathrm{NN}}}=5.02$~TeV using the CoLBT-hydro model with a hybrid hadronization scheme that combines Cooper--Frye particlization, quark coalescence, and string fragmentation. For inclusive charged jets, the calculated in-cone $p/\pi$ and $K/\pi$ ratios are enhanced in Pb+Pb relative to $p+p$ collisions, consistent with the preliminary ALICE measurements. Calculations show that coalescence between shower and thermal partons provides a substantial contribution to these enhancements.

In dijet events, both $p/\pi$ and $K/\pi$ ratios are enhanced around the leading and subleading jets. The enhancements are larger around the subleading jet which on average deposits more energy into the medium, and increase with the cone radius $R$. The $K/\pi$ enhancement is larger at small $R$, whereas the $p/\pi$ enhancement grows more rapidly and dominates at large $R$, reflecting the increasing contribution of the medium response relative to the modified jet shower at larger angles from the jet axis.

After mixed-event acceptance correction and sideband background subtraction, the jet-associated pion, kaon, and proton yields in Pb+Pb collisions all become negative near $|\Delta\phi|\simeq\pi/2$, revealing a depletion consistent with the diffusion wake, whereas the corresponding $p+p$ yields are consistent with zero. The ratios of the deficit magnitudes, $|Y_p|/|Y_\pi|$ and $|Y_K|/|Y_\pi|$, remain finite, are comparable to the Pb+Pb ratios, and exceed the $p+p$ values around the jet peaks. This indicates that the missing hadrons would otherwise have been emitted from jet-excited medium rather than from jet fragmentation. Because these depletion ratios characterize the species composition of the missing yield, they are expected to be less sensitive to the normalization of the background subtraction than the negative yields themselves. Measurements of these ratios, together with statistically significant negative yields for all three species, would therefore provide a new hadrochemical probe of the diffusion wake. Extending these studies to $\gamma$- and $Z$-tagged jets and to light-ion collisions would further clarify the origin of the hadrochemical signatures of jet-induced medium response and provide new constraints on the properties of the QGP.

\textit{Acknowledgments--}
This work is supported in part by National
Natural Science Foundation of China (NSFC) under Grants Nos.~12225503,
12135011, 12547164, 12535010, 12435009 and 11935007, by the Outstanding Leading Talent Team Program of Central China Normal University (XJ2026000302), and by the China Postdoctoral Science Foundation under Grant No. 2025M783388. X.-Y.W was supported in part by the Natural Sciences and Engineering Research Council of Canada  (NSERC) [SAPIN-2026-00047 and SAPIN-2024-00026]. The numerical calculations have been performed on the cluster in the Nuclear Science Computing Center at Central China Normal University (NSC3).

\bibliographystyle{h-physrev5}
\bibliography{references}

%%%%%%%%%%%%%%%%%%
\newpage
\section*{SUPPLEMENTAL MATERIAL}
\noindent This Supplemental Material details the background subtraction procedure for jet-associated yields and the validation of bulk hadrochemistry for the hydrodynamic background.

\subsection{Background Subtraction}
\label{sec:supp_background}
In this section, we provide further details of the background subtraction procedure using the mixed-event and sideband methods. Following the experimental procedure~\cite{CMS:2016qnj}, we calculate the two-dimensional jet-associated hadron yields as functions of $\Delta\eta=\eta_h-\eta_{\rm jet}$ and $\Delta\phi=\phi_h-\phi_{\rm jet}$. To correct for the non-uniform pair acceptance due to the finite pseudorapidity coverage, we employ the mixed-event method,
\begin{equation}
C_i(\Delta\eta,\Delta\phi)
=
\frac{ME_i(0,0)}{ME_i(\Delta\eta,\Delta\phi)}
S_i(\Delta\eta,\Delta\phi),
\label{eq:raw}
\end{equation}
where $i=\pi,K,p$ denotes $\pi^++\pi^-$, $K^++K^-$, and $p+\bar p$, respectively. Here, $S_i$ is the same-event jet--hadron pair distribution, while $ME_i$ is obtained by pairing jets and hadrons from different events,
\begin{align}
S_i(\Delta\eta,\Delta\phi)
&=\frac{1}{N_{\rm jet}}
\frac{d^2N_i^{\rm same}}{d\Delta\eta\,d\Delta\phi},
\nonumber\\
ME_i(\Delta\eta,\Delta\phi)
&=\frac{1}{N_{\rm jet}}
\frac{d^2N_i^{\rm mix}}{d\Delta\eta\,d\Delta\phi},
\label{eq:S-ME}
\end{align}
with $N_{\rm jet}$ being the number of reference jets.

\begin{figure}[t]
\centering
\includegraphics[width=0.5\textwidth]{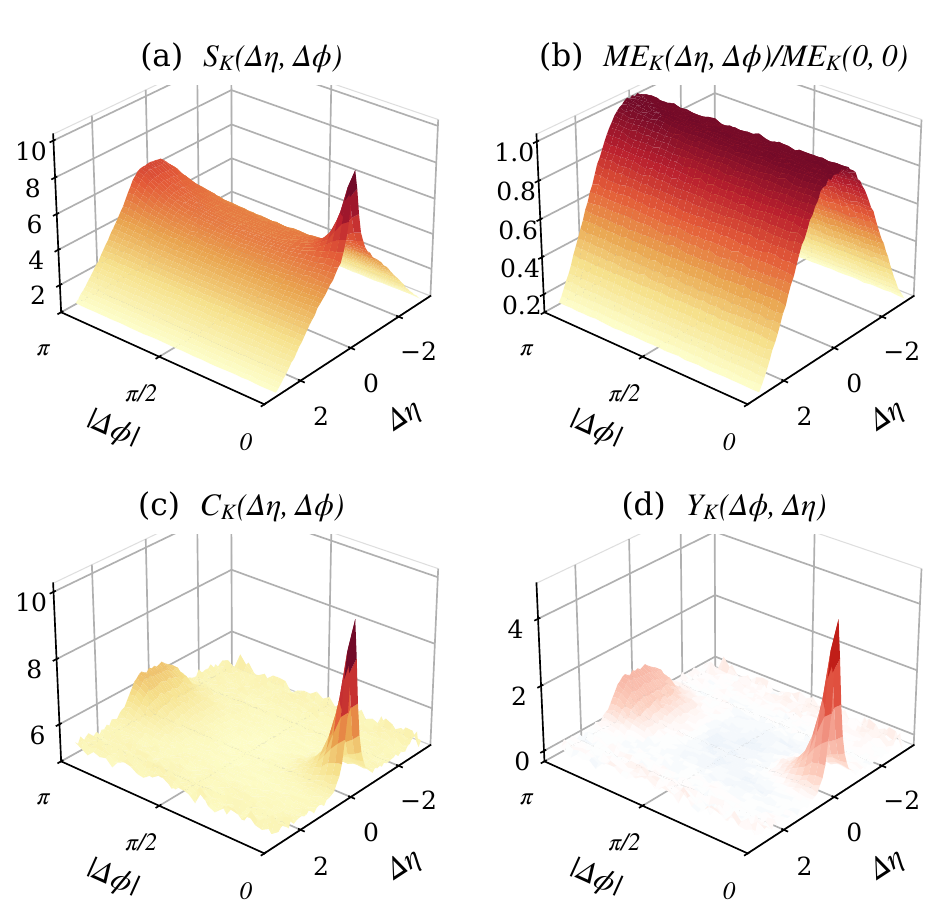}
\caption{Illustration of the mixed-event correction and sideband background subtraction for jet--kaon correlations with $1.5<p_{\mathrm{T}}^h<4$~GeV/$c$. (a) Same-event distribution $S_i(\Delta\eta,\Delta\phi)$. (b) Mixed-event distribution $ME_i(\Delta\eta,\Delta\phi)$. (c) Acceptance-corrected distribution $C_i(\Delta\eta,\Delta\phi)$. (d) Final background-subtracted yield $Y_i(\Delta\eta,\Delta\phi)$.
}
\label{fig:supp_mixing}
\end{figure}

To illustrate this procedure, we show in Fig.~\ref{fig:supp_mixing}(a) the same-event jet--kaon correlation, which contains near- and away-side peaks on top of a broad background. The corresponding mixed-event distribution in Fig.~\ref{fig:supp_mixing}(b) accounts for the pair-acceptance effect, with the correction factor normalized at $\Delta\eta=\Delta\phi=0$. After the mixed-event correction, the acceptance-corrected correlation $C_i(\Delta\eta,\Delta\phi)$ in Fig.~\ref{fig:supp_mixing}(c) includes jet-associated peaks together with the residual background.

%%%%%%%%%%%%%%%%%%
%%%%%%%%%%%%%%%%%%

To remove the residual background, including uncorrelated particle pairs and long-range correlations, we employ the sideband method~\cite{CMS:2016qnj}. The background at each $\Delta\phi$ is estimated from the acceptance-corrected yield averaged over large-$|\Delta\eta|$ sideband region,
\begin{equation}
B_i(\Delta\phi)
=
\frac{1}{\Delta\eta_{\rm SB}}
\int_{\rm SB}d\Delta\eta\,
C_i(\Delta\eta,\Delta\phi),
\label{eq:sideband_background}
\end{equation}
where $\Delta\eta_{\rm SB}$ is the width of the sideband intervals. Assuming that the background is approximately independent of $\Delta\eta$, the final jet-associated hadron yield is obtained as
\begin{equation}
Y_i(\Delta\eta,\Delta\phi)
=
C_i(\Delta\eta,\Delta\phi)-B_i(\Delta\phi).
\label{eq:jet_correlated_yield}
\end{equation}
The resulting kaon distribution is shown in Fig.~\ref{fig:supp_mixing}(d). The final background-subtracted distribution reveals the near-side and away-side jet peaks, as well as a negative yield around $|\Delta\phi|\simeq\pi/2$.

\begin{figure}[tb]
\centering \includegraphics[width=1\columnwidth]{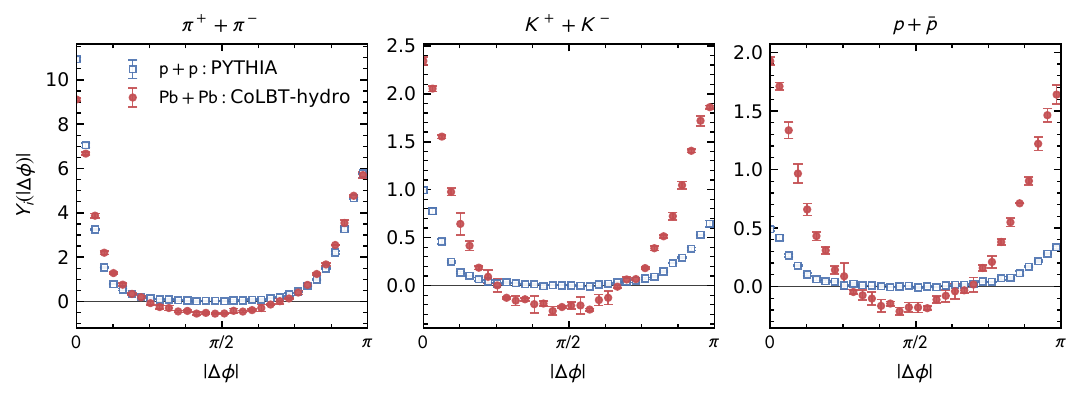}
\caption{Background-subtracted per-jet yields of associated $\pi$ (left), $K$ (middle), and $p$ (right) as functions of $|\Delta\phi|$, integrated over $|\Delta\eta|<1$. Blue open squares show $p+p$ collisions, and red filled circles show 0--10\% Pb+Pb collisions, at $\sqrt{s_{NN}}=5.02$~TeV. The jet and associated-hadron selections are the same as in Figure.~\ref{fig:dijet-2D}.}
\label{fig:dijet-yields-phi} 
\end{figure}

To examine the azimuthal dependence of jet associated identified-hadron yields, we integrate the background-subtracted two-dimensional distributions $Y_i(\Delta\eta,\Delta\phi)$ over $|\Delta\eta|<1$,
\begin{equation}
Y_i(|\Delta\phi|)
=
\frac{1}{2}\int_{-1}^{1} d\Delta\eta\,
\left[
Y_i(\Delta\eta,|\Delta\phi|)
+
Y_i(\Delta\eta,-|\Delta\phi|)
\right].
\label{eq:phi_projection}
\end{equation}
The projected distributions are shown in Fig.~\ref{fig:dijet-yields-phi} where the central Pb+Pb collision results are calculated from for the CoLBT-hydro model and the p+p results are obtained from PYTHIA. 
Around two jet peaks, the kaon and proton yields are enhanced in Pb+Pb relative to $p+p$ collisions, while the pion yield is slightly suppressed on the near side and remains comparable on the away side. The enhanced kaon and proton yields support the role of coalescence and contribute to the increased $p/\pi$ and $K/\pi$ ratios discussed in the main text.
Near $|\Delta\phi|\simeq\pi/2$, all three hadron species (pions, kaons, and protons) exhibit negative yields in Pb+Pb collisions, whereas the corresponding $p+p$ yields are consistent with zero.

%%%%%%%%%%%%%%%%%%
%%%%%%%%%%%%%%%%%%
%%%%%%%%%%%%%%%%%%
\begin{figure}[tb]
\centering \includegraphics[width=1\columnwidth]{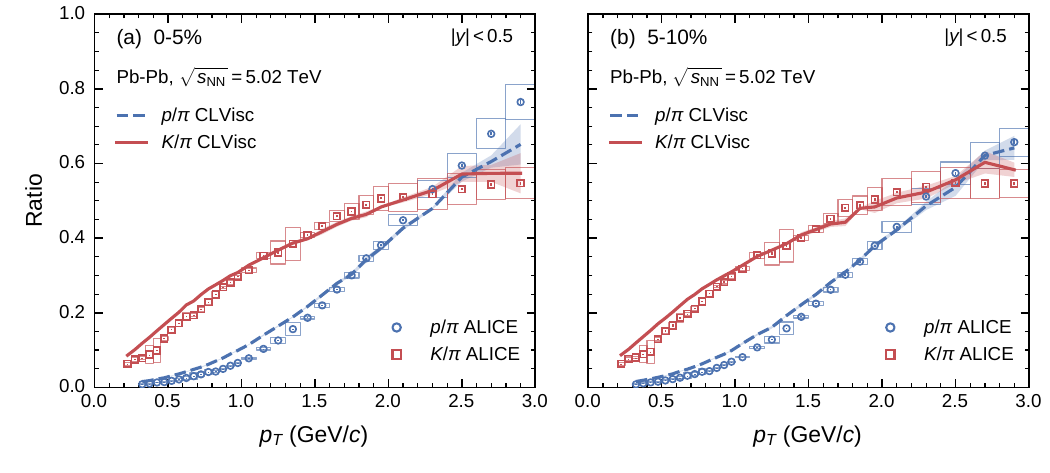}
\caption{Transverse-momentum dependence of $(p+\bar p)/(\pi^++\pi^-)$ (blue circles) and $(K^++K^-)/(\pi^++\pi^-)$ (red squares) ratios at $|y|<0.5$ in (a) 0--5\% and (b) 5--10\% Pb+Pb collisions at $\sqrt{s_{\mathrm{NN}}}=5.02$~TeV. Curves show CLVisc calculations and symbols show the ALICE measurements~\cite{ALICE:2019hno}.}
\label{fig:hydro-ratio} 
\end{figure}
%%%%%%%%%%%%%%%%%%

\subsection{Bulk hadrochemistry}
\label{sec:supp_hydro_ratio}

The hadrochemical signatures of jet-induced medium response discussed in the main text are defined on top of a thermal background. 
In the CoLBT-hydro calculation, both the Cooper--Frye emission from the freeze-out hypersurface and the supply of thermal partons for shower--thermal coalescence are determined by the hydrodynamic evolution. 
Therefore, the interpretation of dijet results requires that the pure hydrodynamic simulations reproduce the hadrochemistry of the bulk medium. 
To validate this, Fig.~\ref{fig:hydro-ratio} compares the $(p+\bar p)/(\pi^++\pi^-)$ and $(K^++K^-)/(\pi^++\pi^-)$ ratios as calculated from CLVisc hydrodynamics with the ALICE measurements~\cite{ALICE:2019hno} at $|y|<0.5$ in 0--5\% and 5--10\% Pb+Pb collisions at $\sqrt{s_{\mathrm{NN}}}=5.02$~TeV. Here, the hydrodynamic evolution uses the same parameters as in the CoLBT-hydro calculation, but without jets.
Hadrons are produced on the freeze-out hypersurface at $T_f=150$~MeV through the Cooper--Frye prescription and then undergo resonance decays in SMASH, consistent with the treatment of jet-associated hadrons. 

The hydrodynamics calculation captures the characteristic momentum dependence of both $p/\pi$ and $K/\pi$ ratios. At low $p_{\mathrm{T}}$, the ratios are governed by hadron masses and the freeze-out temperature, whereas at intermediate $p_{\mathrm{T}}$, $p/\pi$ ratio rises and develops a maximum while $K/\pi$ ratio rises more gently. This pattern reflects the mass ordering effect induced by the radial flow of the medium, which shifts heavier hadrons to higher $p_{\mathrm{T}}$. A good agreement is found in both 0--5\% and 5--10\% centrality classes, supporting the use of this hydrodynamic background for 0--10\% events in the dijet analysis presented in this work. Note that at higher $p_{\mathrm{T}}$, a pure hydrodynamic description is expected to break down since hadron production becomes more dominated by coalescence and fragmentation. In the CoLBT-hydro calculation, these contributions are supplied by the shower partons of the quenched jets, which are not included in the pure hydrodynamic calculation. Therefore, the comparison with the ALICE data shown here is restricted to $p_{\mathrm{T}}\lesssim 3$~GeV/$c$, where the hydrodynamic component is expected to dominate.

\end{document}